\documentstyle[eqsecnum,epsfig, aps,prd]{revtex}

\begin{document}
\draft
\preprint{hep-ph/0202016}
\title{ Loop effects and non-decoupling property of SUSY QCD 
         in $g b\rightarrow tH^{-}$ }
\author{Guangping Gao $^{a,b}$, Gongru Lu $^a$, 
        Zhaohua Xiong $^{b,c}$  and Jin Min Yang $^b$}
\address{$^a$ Physics Department, Henan Normal University, 
         Xinxiang 453002, China}
\address{$^b$ Institute of Theoretical Physics, Academia Sinica, 
           Beijing 100080, China} 
\address{$^c$ Institute of High Energy Physics, Academia Sinica,
         Beijing 100039, China}
\date{\today}
\maketitle

\begin{abstract}
One-loop SUSY QCD radiative correction to $gb \rightarrow tH^{-}$ 
cross section is calculated in the Minimal Supersymmetric Standard
Model. We found that SUSY QCD is non-decoupling if the gluino mass
and the parameter $\mu$, $A_t$ or $A_b$ are at the same order and
get large. The non-decoupling contribution can be enhanced by large 
$\tan\beta$ and therefore large corrections to the hadronic production 
rates at the Tevatron and LHC are expected in the large $\tan\beta$ 
limit. The fundamental reason for such non-decoupling behavior is 
found to be some couplings in the loops being  proportional to SUSY 
mass parameters. 

\end{abstract} 


\section{Introduction}
\label{sec:sec1}

Although the Standard Model (SM) is phenomenologically successful,
it is arguably an effective theory and new physics must exist at
high energy scales. Among all elementary particles predicted by the SM,
top quark and Higgs boson may hold the key to new physics since they are 
most related to the electroweak symmetry breaking.
An intensive study of the properties of top quark and Higgs 
boson will be one of the primary tasks of particle physics in the new 
millennium.

So far the most intensively studied new physics model is the Minimal 
Supersymmetric Standard Model (MSSM)~\cite{HaberKane}. This model
predicts the existence of five Higgs bosons, $H^0$, $h^0$, $A^0$ 
and $H^{\pm}$, all of which couples to top quark. Compared to the  
couplings in the SM, the coupling $tbH^-$  is an utterly new coupling. 
Studies \cite{tbh-susy} show that this coupling is sensitive to 
quantum corrections and may be a good probe of the MSSM.
Although this coupling could be measured from top quark decay process
$t\to H^+ b$ if the charged Higgs is sufficiently light, the direct production 
of a top quark associated with the charged Higgs boson through the subprocess 
$gb\to tH^-$ at hadron colliders will be a good probe for  $tbH^-$ 
coupling \cite{gbth}.   In this work we calculate the one-loop SUSY QCD 
corrections to this process owing to the following motivations. 
Firstly, if the charged Higgs boson is heavy, $m_{H^+}>m_t+m_b$, 
as a main charged Higgs production channel\cite{Bawa90}, the process 
$gb\to tH^-$ will provide sizable cross section at the Tevatron and LHC. 
The supersymmetric radiative corrections, especially the SUSY-QCD corrections, 
to this high energy process may be significant, as was found for other 
similar processes \cite{Guasch95,Carena00,hbb,hpt,Dobado02}.  
Secondly, some recent studies \cite{hbb,hpt,Dobado02} 
showed that SUSY-QCD may be non-decoupling in some processes
involving Higgs bosons. As is well known, the 
decoupling theorem \cite{TAC} states that under certain conditions in a given 
quantum field theory with light and heavy particles, if the heavy particles 
are integrated out to all orders in perturbation theory, the remaining 
effective action to be valid at energies much lower than the heavy particle 
masses does not show any trace of these heavy particles. If SUSY QCD is
non-decoupling in some cases, we need a proper understanding and
thus we need to further investigate such non-decoupling property of SUSY QCD.  
$gb\to tH^-$  will be an ideal process for this purpose.

This paper is organized as follows. 
In Section~\ref{sec:calculations}  we present the formula for the one-loop 
SUSY-QCD corrections to the $gb \rightarrow tH^{\pm}$ process. 
In Section~\ref{sec:parameters} we  scan the parameter space of MSSM
to estimate the size of SUSY-QCD corrections.
In Section~\ref{sec:decouple} we study the decoupling behavior of SUSY QCD. 
A discussion on how the decoupling and non-decoupling take place is also given.
Finally, the conclusions are summarized in Section~\ref{sec:conclusions}.  

\section{Calculations}
\label{sec:calculations}
\begin{figure}
\begin{center}
\epsfig{file=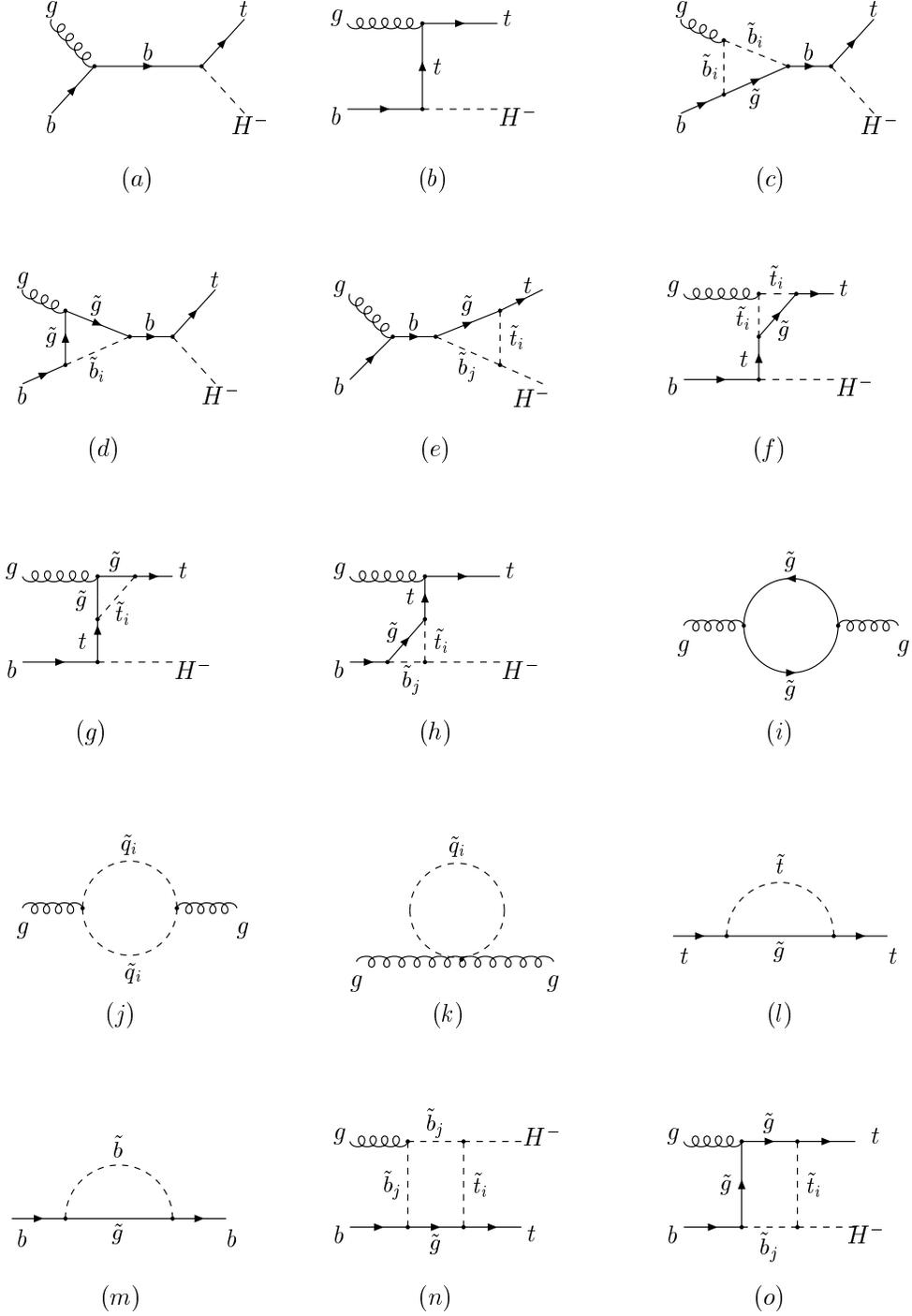,width=14cm}
\caption{ Feynman diagrams of $gb\rightarrow tH^{-}$ with one-loop SUSY QCD corrections: 
$(a)$ and $(b)$ are tree level diagrams ; $(c)-(e)$ are one-loop vertex diagrams for s-channel ;
$(f)-(h)$ are one-loop vertex diagrams for t-channel ; $(i)-(m)$ are
self-energy diagrams; $(n,o)$ are the box diagrams.}
\label{fig:feyman}
\end{center}
\end{figure}

The subprocess  $gb\rightarrow tH^-$ occurs through both s-channel and t-channel.
The tree-level amplitude is given by 
\begin{equation}
M_{0}=M_{0}^{(s)}+M_{0}^{(t)},
\end{equation}
where $M_{0}^{(s)}$ and $M_{0}^{(t)}$ represent 
the amplitudes arising from the s-channel diagram shown in Fig.~\ref{fig:feyman}$(a)$ 
and  the t-channel diagram shown in Fig.~\ref{fig:feyman}$(b)$, respectively. 
Their amplitudes can be expressed as 
\begin{eqnarray}
M_{0}^{(s)}&=&\frac{igg_{s}V_{tb}}{\sqrt{2}m_{W}
(\hat{s}-m_{b}^{2})}\overline{u}(p_{t})\left[2\eta_t p_{b}^{\mu}P_L
+2\eta_bp_{b}^{\mu}P_{R} -\eta_t
\gamma^{\mu}{\not{k}}P_{L}
-\eta_b\gamma^{\mu}{\not{k}} P_{R}\right]u(p_{b})\varepsilon_{\mu}(k)
T_{ij}^{a}, \\
M_{0}^{(t)}&=& \frac{igg_{s}V_{tb}}{\sqrt{2}m_{W}(\hat{t}
-m_{t}^{2})}\overline{u}(p_{t})
\left[2\eta_tp_{t}^{\mu}P_{L}+2\eta_bp_{t}^{\mu}P_{R}
-\eta_t\gamma^{\mu}{\not{k}} P_{L}
-\eta_b\gamma^{\mu}{\not{k}}P_{R}\right]
u(p_{b})\varepsilon_{\mu}(k)T_{ij}^{a},
\end{eqnarray}
where $P_{R,L}\equiv (1\pm \gamma_5)/2$, and  
$p_t$, $p_b$ and $k$ are the momenta of the outgoing top quark, 
the incoming bottom quark and the incoming gluon, respectively.    
$\hat{s}$ and $\hat{t}$ are the subprocess Mandelstam variables defined by 
$\hat{s}=(p_{b}+k)^2=(p_t+p_{H^-})^2$ and $\hat{t}=(p_t-k)^2=(p_{H^-}-p_b)^2$.
$T^{a}$ are the $SU(3)$ color matrices and $\tan\beta=v_2/v_1$ is the ratio of 
the vacuum expectation values of the two Higgs doublets. The constants $\eta_{b,t}$
are defined by $\eta_b= m_b\tan\beta$ and $\eta_t=m_{t}\cot\beta$.

The one-loop Feynman diagrams of SUSY QCD corrections 
are shown in  Fig.~\ref{fig:feyman}$(c)$-$(o)$.
In our calculations we use dimensional regularization to 
control all the ultraviolet divergences in the virtual loop corrections 
and we adopt the on-mass-shell renormalization scheme.
The renormalization condition for the coupling constant $g_s$ 
is similar to that for the coupling constant $e$ in QED, 
i.e., the coupling of the photon (gluon) 
to a pair of fermions is required to recover the tree-level  
result in the limit of zero momentum transfer. This condition 
yields 
\begin{eqnarray} \label{gs}
\frac{\delta g_s}{g_s}=-\frac{1}{2} \delta Z_2^g,
\end{eqnarray}    
where $\delta g_s$ and  $ Z_2^g$ are the renormalization constants
defined by $g^0_s \equiv g_s+\delta g_s$ and 
$A^0_{\mu} \equiv \sqrt{Z_2^g} A_{\mu}$ with $ g^0_s$ denoting 
bare coupling constant and $A^0_{\mu}$ bare gluon fields (color
index suppressed).  
  
Including the one-loop SUSY QCD corrections, the renormalized 
amplitude for $gb\rightarrow tH^{-}$ can be written as
\begin{eqnarray}
M_{ren}&=& M_{0}^{(s)} +M_{0}^{(t)} +\delta M,
\end{eqnarray}
where $\delta M$ represents the one-loop SUSY QCD corrections
given by
\begin{eqnarray}
\delta M =\delta M^{V_{1}(s)} +\delta M^{V_{2}(s)}
+\delta M^{s(s)}+\delta M^{box}
+\delta M^{V_{1}(t)} +\delta M^{V_{2}(t)}+\delta M^{s(t)} .
\end{eqnarray}
Here $\delta M^{V_1(s)}$, $\delta M^{V_2(s)}$ and  $\delta M^{s(s)}$ 
represent the renormalized vertex $gb\bar{b}$, 
$t \bar b H^-$ and the renormalized propagator in the s-channel diagram, 
respectively. Similar definitions are for 
$\delta M^{V_1(t)}$, $\delta M^{V_2(t)}$ and 
$\delta M^{s(t)}$ in the t-channel diagram.
The contribution of the box diagram is denoted by $\delta M^{box}$.
Each $\delta M^{l}$ can be decomposed as 
\begin{eqnarray}
\delta M^{l}&=& \frac{igg_{s}^{3} T_{ij}^{a}V_{tb}}
{16\sqrt{2}\pi^{2}m_{W}} C^{l}\overline{u}
(p_{t})\left\{F_{1}^{l} \gamma^{\mu}P_{L} +F_{2}^{l} \gamma^{\mu}P_{R}
+F_{3}^{l}p_{b}^{\mu}P_{L} +F_{4}^{l}p_{b}^{\mu}P_{R}
+F_{5}^{l}p_{t}^{\mu}P_{L} +F_{6}^{l}p_{t}^{\mu}P_{R} 
\right.\nonumber \\ 
& &\left. +F_{7}^{l}\gamma^{\mu}{\not{k}}P_{L}
+F_{8}^{l} \gamma^{\mu}{\not{k}}P_{R}
+F_{9}^{l}p_{b}^{\mu}{\not{k}}P_{L}
+F_{10}^{l}p_{b}^{\mu}{\not{k}}P_{R}
+F_{11}^{l}p_{t}^{\mu}{\not{k}}P_{L} 
+F_{12}^{l}p_{t}^{\mu}{\not{k}}P_{R}\right\}
u(p_{b})\varepsilon_{\mu}(k),
\end{eqnarray}
where the  coefficients  $C^{l}$ and the form factors 
$F_{n}^{l}$ are given explicitly in Appendix A and B, 
respectively . 
We have checked that all the ultraviolet divergences  
canceled as a result of renormalizability of MSSM.

The amplitude squared is given by
\begin{eqnarray}
\overline{\sum}\left|M_{ren}\right|^{2} 
=\overline{\sum}\left|M_{0}^{(s)}+M_{0}^{(t)}\right|^{2}
+2Re\overline{\sum}\left[
\left(M_{0}^{(s)}+M_{0}^{(t)}\right)^{\dag}\delta M\right],
\end{eqnarray}
where
\begin{eqnarray} 
\overline{\sum}\left|M_{0}^{(s)} +M_{0}^{(t)}\right|^{2} &=&
\frac{2g^{2}g_{s}^{2}|V_{tb}|^2}{N_{C}m_{W}^{2}}
\left\{\frac{1}{(\hat{s}-m_{b}^{2})^{2}}
\left[(\eta_b^2+\eta_t^2)
( p_{b}\cdot k p_{t}\cdot k +2p_{b}\cdot kp_{b}\cdot p_t
-m_{b}^{2}p_{t}\cdot k p_{t}-m_{b}^{2}p_{b}\cdot p_{t})
\right.\right. \nonumber \\
& &\left.\left. +2m_{b}^{2}m_{t}^{2}(p_{b}\cdot k
- m_{b}^{2})\right]
+\frac{1}{(\hat{t}-m_{t}^{2})^{2}}\left[(\eta_b^2+\eta_t^2)(p_{b}\cdot kp_{t}\cdot k
+m_{t}^{2}p_{b}\cdot k-m_{t}^{2}p_{b}\cdot p_{t})
\right.\right. \nonumber \\ 
& &\left. \left.+2m_{b}^{2}m_{t}^{2}(p_{t}\cdot k -m_{t}^{2})\right]
+\frac{1}{(\hat{s}-m_{b}^{2})(\hat{t}-m_{t}^{2})}
\left[(\eta_b^2+\eta_t^2)(2p_{b}\cdot kp_{t}\cdot k
 +2p_{b}\cdot kp_{b}\cdot p_{t} \right.\right. \nonumber\\
 & &\left.\left.-2(p_{b}\cdot p_{t})^{2}
-m_{b}^{2}p_{t}\cdot k  +m_{t}^{2}p_{b}\cdot k)
 +2m_{b}^{2}m_{t}^{2}(p_{t}\cdot k -p_{b}\cdot k -2p_{b}\cdot p_{t})\right]\right\},
\end{eqnarray}
\begin{eqnarray}
\label{eq29}
\overline{\sum}\left(M_{0}^{(s)}+M_{0}^{(t)}\right)^{\dag}\delta M
=-\frac{g^{2}g_{s}^4|V_{tb}|^2} {64N_C\pi^2m_W^2}
\sum_{n=1}^{12}\left[\frac{1}{\hat{s}-m_b^{2}}h_{n}^{(s)}
+\frac{1}{\hat{t}-m_{t}^{2}}h_{n}^{(t)}\right]C^lF_{n}^{l}.
\end{eqnarray}
Here the color factor $N_{C}=3$, and $h_{n}^{(s)}$ and
$h_{n}^{(t)}$ can be found in Appendix A.

\vspace{0.1cm}

The cross section for the parton process $gb\rightarrow tH^{-}$ is
\begin{equation}
\hat{\sigma}(\hat s) =\int_{\hat{t}_{min}}^{\hat{t}_{max}}\frac{1}{16\pi
\hat{s}^2} \overline{\Sigma}|M_{ren}|^{2}d\hat{t}\,,
\end{equation}
with
\begin{eqnarray}
\hat{t}_{max,min} =\frac{1}{2}\left\{m_{t}^{2} +m_{H^{-}}^{2} -\hat{s}
\pm \sqrt{[\hat{s} -(m_{t}+m_{H^{-}})^{2}][\hat{s} -(m_{t} -m_{H^{-}})^{2}]}
\right\}.
\end{eqnarray}

The total hadronic cross section for $pp({\rm or}~p\bar{p}) \rightarrow
tH^{-}+X$ can be obtained by folding the subprocess cross section
$\hat{\sigma}$ with the parton luminosity
\begin{equation}
\sigma(s)=\int_{\tau_0}^1 \!d\tau\, 
\frac{dL}{d\tau}\, \hat\sigma (\hat s=s\tau) ,
\end{equation}
where $\tau_0=(m_t+m_{H^-})^2/s$, and $s$ is the $p 
p({\rm or}~p\bar{p})$ 
center-of-mass energy squared. $dL/d\tau$ is the parton luminosity given by 
\begin{equation}
\frac{dL}{d\tau}=\int^1_{\tau} \frac{dx}{x}[f^p_g(x,Q)
f^{p}_b(\tau/x,Q)+(g\leftrightarrow b)],
\end{equation}
where $f^p_b$ and $f^p_g$ are the bottom
quark and gluon distribution functions in a proton, respectively. 
In our numerical calculation, we use the CTEQ5L parton 
distribution functions~\cite{pm} with $Q=m_t+m_{H^-}$.

To show the size of the corrections, we define the relative
quantity as
\begin{equation}
\Delta_{SQCD}=\frac{\sigma-\sigma_0}{\sigma_0},
\end{equation}
where $\sigma_0$ is the tree-level cross section.

\section{Numerical results}  
\label{sec:parameters}

Before performing numerical calculations, we take a look at the
relevant parameters involved. 

For the SM parameters, we took 
$m_W=80.448$ GeV, $m_Z=91.187$ GeV, $m_t=176$ GeV, $m_b=4.5$ GeV, 
and used the two-loop running coupling constant $\alpha_s(Q)$. 

For the SUSY parameters, apart from the charged Higgs mass, gluino mass 
and $\tan\beta$, the mass parameters of stops and sbottoms are involved.
The mass square matrices of stop and sbottoms take the form
($q=t$ or $b$)~\cite{susyint}
\begin{equation}
M_{\tilde q}^2 =\left(\begin{array}{cc}
m_{{\tilde q}_L}^2& m_qX_q^\dag\\
 m_qX_q& m_{{\tilde q}_R}^2
    \end{array} \right), 
\end{equation}
where  
\begin{eqnarray}
m_{{\tilde q}_L}^2 &=& m_{\tilde Q}^2+m_q^2-m_Z^2(\frac{1}{2}
+e_q\sin^2\theta_W)\cos(2\beta),\nonumber\\ 
m_{{\tilde q}_R}^2 &=& m_{\tilde U,\tilde D}^2+m_q^2
+e_qm_Z^2 \sin^2\theta_W\cos(2\beta),\nonumber\\
X_q&=& \left\{ \begin{array}{ll} A_t-\mu\cot\beta, & {\rm for~} q=t, \\ 
                                 A_b-\mu\tan\beta, & {\rm for~} q=b.
               \end{array} \right. 
\label{smass1}
\end{eqnarray}
Here $m_{\tilde Q}^2$, $m_{\tilde{U}}$ and $m_{\tilde{D}}^2$ are 
soft-breaking mass terms for left-handed squark doublet 
$\tilde Q$, right-handed up squark $\tilde U$ and down squark $\tilde D$, 
respectively. 
$A_b (A_t)$  is the coefficient of the trilinear term 
$H_1 \tilde Q \tilde D$ ($H_2 \tilde Q \tilde U$ )
in soft-breaking terms and $\mu$ the bilinear coupling of the 
two Higgs doublet in the superpotential. 
Thus the SUSY parameters involved in stop and sbottom mass
matrices are  
\begin{eqnarray*} 
m_{\tilde{Q}}, m_{\tilde{U}}, m_{\tilde{D}}, A_t, A_b, \mu, \tan\beta.  
\end{eqnarray*}
The mass square matrices are  diagonalized by  unitary rotations  ($q=t$ or $b$) 
\begin{eqnarray} 
            R^q=\left (
             \begin{array}{cc}
            \cos\theta_q       &\sin\theta_q\\
           -\sin\theta_q       &\cos\theta_q\\
           \end{array} \right )
\end{eqnarray}
which relates the weak-eigenstates $({\tilde q}_L,{\tilde q}_R)$ 
to the mass eigenstates  $({\tilde q}_1,{\tilde q}_2)$.    
Then the stop and sbottom masses as well as the mixing angles are 
obtained by 
\begin{eqnarray} 
m_{\tilde{q}_{1,2}}&=&\frac{1}{2}\left[
m_{{\tilde q}_L}^2+m_{{\tilde q}_R}^2
\pm\sqrt{\left(m_{{\tilde q}_L}^2-m_{{\tilde q}_R}^2\right)^2
+4m_q^2X_q^2}\right],\nonumber\\
\tan2\theta_{q} &=& \frac{2m_qX_q}
{m_{{\tilde q}_L}^2-m_{{\tilde q}_R}^2}.
\label{smass2}
\end{eqnarray}
To determinate the mixing angles completely,  
we adopt the convention in \cite{Djouadi98} which sets $\theta_{q}=\pi/4$
if $m_{{\tilde q}_L}=m_{{\tilde q}_R}$ and 
shifts $\pi/2$ to $\theta_{q}$ if 
$m_{{\tilde q}_L}>m_{{\tilde q}_R}$. Thus $\theta_{q}$ lies in the range 
$-\frac{\pi}{4}\leq\theta_{q}\leq \frac{3}{4}\pi$. 

To find out the size of the one-loop SUSY QCD effects,
we performed a scan over the  nine-dimensional 
parameter space:  $m_{\tilde{Q}}$, $m_{\tilde{U}}$, $m_{\tilde{D}}$,
$A_t$, $A_b$, $\mu$, $tan\beta$, $m_{H^-}$, $m_{\tilde{g}}$. 
In scanning we restricted $m_{H^-}$, $A_t$,  $A_b$ and $\mu$ to the sub-TeV region
and required $m_{H^-}>150$ GeV. Other mass parameters are assumed to be smaller than 5~TeV. 
In addition, we consider the following experimental constraints: 
\begin{itemize}
\item[{\rm(1)}]
   $\mu>0$ and a large $\tan\beta$ in the range $5\le\tan\beta\le 50$, which might be 
   favored by the recent muon $g-2$ measurement~\cite{Brown01}. 
\item[{\rm(2)}]
   The LEP and CDF lower mass bounds on gluino, stop and sbottom~\cite{PDG00}
\begin{eqnarray}
m_{{\tilde t}_1}\geq 86.4~GeV,~
m_{{\tilde b}_1}\geq 75.0~GeV,~ m_{\tilde{g}}\geq 190~GeV .
\end{eqnarray}
\end{itemize}

\begin{figure}
\begin{center}
\epsfig{file=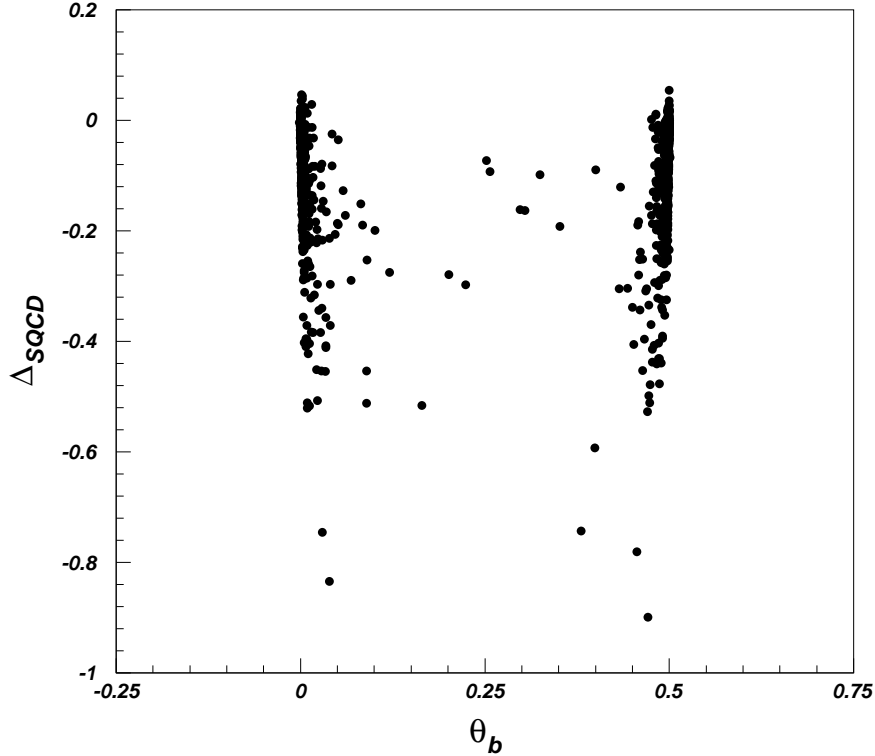,width=12cm}
\caption{The scatter plot in the plane of  $\Delta_{SQCD}$ versus  $\theta_b$.  
         The scan was performed over nine SUSY parameters. $\theta_b$ is in unit of $\pi$.} 
\label{fig:scan}
\end{center}
\end{figure}

The scan results are plotted in the plane of $\Delta_{SQCD}$ versus $\theta_b$ in Fig.~\ref{fig:scan}.  
From Fig.~\ref{fig:scan} one can see that in most parameter space the mixing of sbottoms is small
while the one-loop SUSY QCD effect can be quite large. In some part of the parameter space,  the
correction size can be larger than 20\% which cannot be neglected in the study of this process
at LHC.

\section{Decoupling property of SUSY QCD}
\label{sec:decouple}
To find out if SUSY QCD is decoupling from the process
$gb \rightarrow tH^-$ in the large limit of SUSY mass 
parameters, we fix the charged Higgs mass as $m_{H^-}
=250~GeV$ and consider the following scenarios.

(1) {\em Scenario A:}~~
All squark (collectively denoted by $m_S$), 
gluino masses and $\mu$ or $A$ parameters are of the same size 
and much heavier than the electroweak scale, i.e.,

\begin{eqnarray}
m_S \sim m_{\tilde Q} \sim m_{\tilde U} \sim m_{\tilde D} 
\sim m_{\tilde g}\sim \mu~ or~ A_t ~or ~A_b \sim m_S  \gg M_{EW}
\end{eqnarray} 
In this case, both mixings in the sbottom and stop sectors
reach their maximal values, i.e., $\theta_t\sim\pm\frac{\pi}{4}$,
$\theta_b\sim\pm\frac{\pi}{4}$. As shown in 
Eqs.~(\ref{coupling}-\ref{coupling2}), 
the couplings $\alpha_{ij}$ in the vertex 
$H^{-}\tilde{t}_{i}\tilde{b}_{j}$ are proportional to the 
linear combination of $\mu+A_b \tan\beta$ and 
$\mu+A_t \cot\beta$. Considering that the couplings $\alpha_{ij}$ 
are proportional to $m_S$ as $\mu$ or $A_{t,b}$ gets large as 
$m_S$, and the loop scalar integral functions $C_0$ 
goes to $-1/2m_S^2$ as $m_S\gg \hat{s}$ (see Eq.~(\ref{c0a}) and 
Eq.~(B8) in Ref.~\cite{hpt}), one can infer that the terms 
$\alpha_{ij}A_{ij}^{2(L,R)}C^6_{0}m_{\tilde{g}}$
which arise from the vertex correction to $H^{-}tb$ 
do not vanish but go to a non-zero constant, showing a 
clear indication of {\em non-decoupling} behavior.
In fact, such non-decoupling behavior will happen 
as long as the gluino mass and $\mu$ or $A$ 
parameters are of the same order, not necessarily 
degenerate. Actually, from the expansions of the three- 
and four-point loop integrals in the asymptotic large 
mass limit in Appendix C, one can see this fact. 

\begin{figure}[htb]
\begin{center}
\epsfig{file=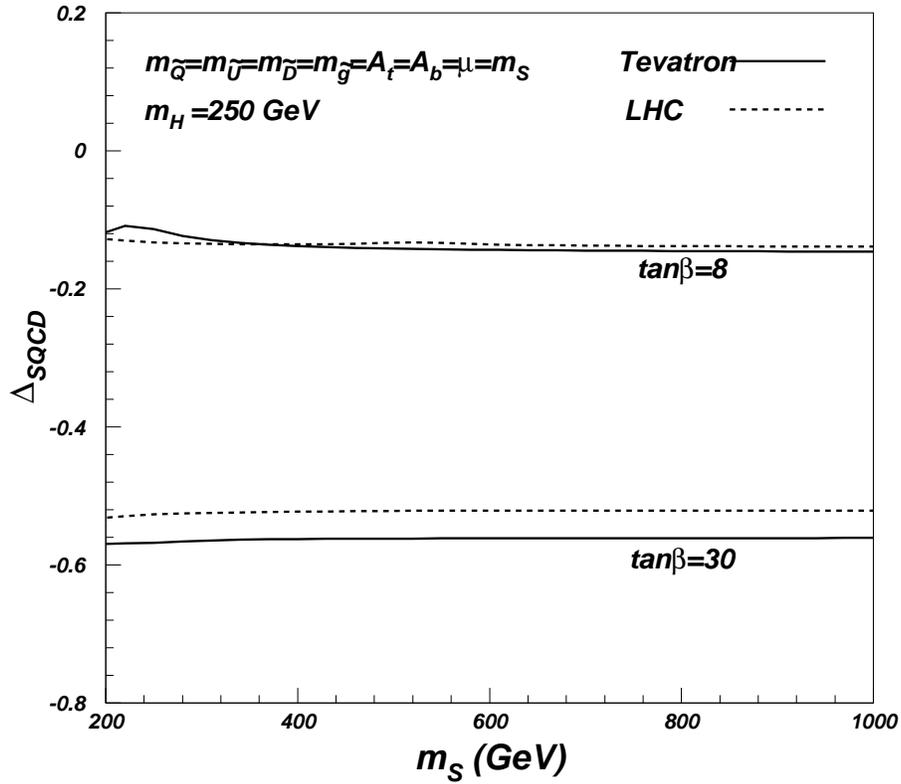,width=12cm}
\caption{ Non-decoupling behavior of $\Delta_{SQCD}$ with
$m_{\tilde Q} = m_{\tilde U} = m_{\tilde D} = m_{\tilde g} = A_b = A_t
= \mu = m_S$ and for different values of $\tan\beta$.
Corrections at the Tevatron with $\sqrt{s} = 2~TeV$ (solid lines)   
and at the LHC $\sqrt{s} = 14~TeV$ (dashed lines) are plotted  
respectively.}
\label{fig:sigma-gq}
\end{center}
\end{figure}

As illustrative examples, we plot the dependence of the 
SUSY-QCD correction to $gb\rightarrow tH^{-}$ on the common 
SUSY parameter $m_S$ in Fig.~\ref{fig:sigma-gq}. From this figure 
one can see that the non-decoupling behavior indeed happens.
As for the dependence of the non-decoupling effects on $\tan\beta$, 
it is quite involved and complicated. For the parameter values 
chosen in our numerical examples, the corrections are
enhanced by $\tan \beta$. 
   
(2) {\em Scenario B:}~~ The gluino mass and $\mu$ are  of the same order
    (collectively denoted by $m_S$) and get much larger than 
    squark masses and electroweak scale, i.e., 
\begin{eqnarray}
\sim m_{\tilde g}\sim \mu \gg m_S \sim m_{\tilde Q}\sim m_{\tilde U} 
\sim m_{\tilde D}\geq M_{EW} 
\end{eqnarray} 

To keep stop and sbottom masses from getting large, we can set 
$A_t\simeq \mu\cot\beta$, $A_b \simeq \mu\tan\beta$. In this scenario, 
no mixings occur in the sbottom and stop sectors. Apart from the vertex 
correction discussed in {\em Scenario A}, the terms such as 
$\alpha_{ij}A_{ij}^{2(L,R)}D^2_{0}m_{\tilde{g}}$ 
in the box contribution (see Eq.~(\ref{vbox}) and (\ref{d0a})) 
do not vanish either since the four-point integral functions 
$D^2_{(0,1l)}\to 1/m_{\tilde g}^2\ (l=1,2,3)$ 
when $m_{\tilde g}^2\gg\hat{s}$. So in this case the SUSY QCD is 
non-decoupling. This differs from {\em Scenario A} where 
$D_{(0,1l)}\to 1/m_{\tilde{g}}^4$.

Although the non-decoupling effects can arise from more diagrams in 
this case, the reason is the same as in Scenario A, i.e.,  
the couplings $H^{-}\tilde{t}_{i}\tilde{b}_{j}$ are proportional to $\mu$. 
To prove this point fully,  now let's focus our attention on the terms 
arising from the corrections to the $gb\bar{b}$ and $g\tilde{g}\tilde{g}$ 
vertices which are also likely to give the contributions to non-decoupling 
effects in both scenarios we discussed. Firstly, for example we consider 
the term 
\begin{equation}
-\frac{2}{3}\eta_tC^1_{24}
-3\eta_t\left[m_{\tilde{g}}^2C^2_{0}-2C^2_{24}+1/2\right]
+\frac{16\pi^2}{g_{s}^2}\eta_t2\delta Z_{R}^{b}
\end{equation}
in $F_3^{1(s)}$.  From Eq.~(\ref{vselfa}), (\ref{b1}),(\ref{c0a}),(\ref{c24a}),
(\ref{c0b}) and (\ref{c24b}), we draw a conclusion that 
the term indeed cancels out. Further, it is easy to find that all terms 
related with $A^2_{bi}$ are zero in the asymptotic large mass limit.   
This is also valid in the {\em Scenario C} and {\em Scenario D} 
we will study. Secondly, we exam the terms such as 
$\alpha_{ij}A_{ij}^{1(L,R)}D^1_{0}m^2_{\tilde{g}}$ in Eq.~(\ref{vbox}). 
The asymptotic form of the function $D^1_{0}$ in large mass limit is
always proportional to $\frac{1}{m^4_{\tilde{g}}}$ and different from 
$D^2_{0}$ which is proportional to $\frac{1}{m^2_{\tilde{g}}}$ in 
the {\em Scenario B}. So the terms don't cause non-decoupling either.

(3) {\em Scenario C:}~~ Only the gluino mass is very large than other 
SUSY parameters and the electroweak scale

\begin{figure}[hbt]
\begin{center}
\epsfig{file=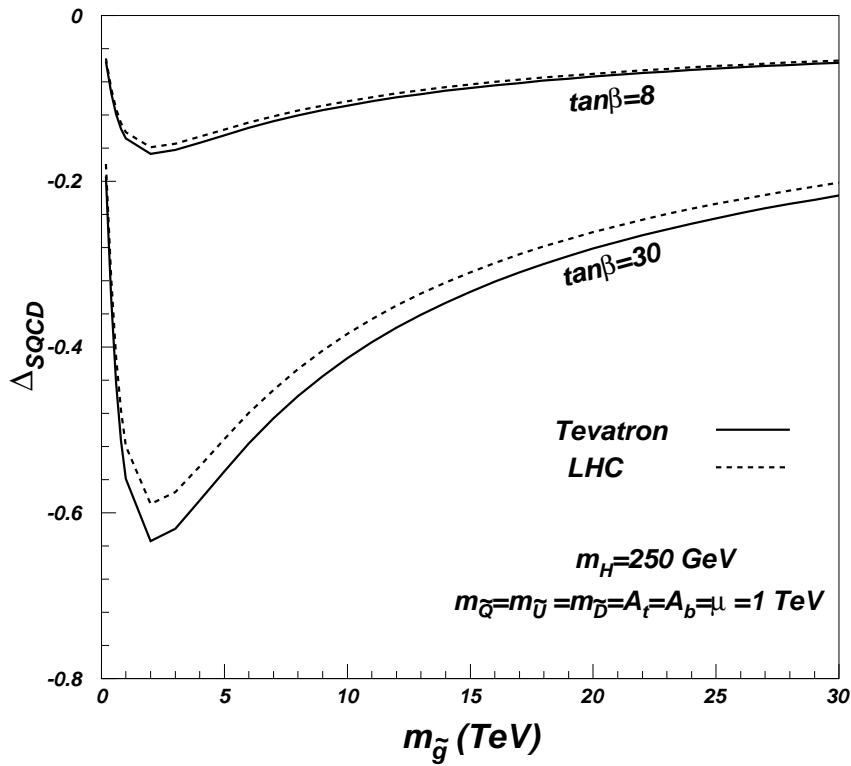,width=12cm}
\caption{Behavior of $\Delta_{SQCD}$ in the large $m_{\tilde g}$
limit with fixed $m_{\tilde Q} = m_{\tilde U} = m_{\tilde D} 
= A_b = A_t = \mu = 1$ TeV and for different values of 
$\tan \beta$. The solid and dashed  
lines correspond to corrections at the Tevatron with $\sqrt{s} = 2 TeV$  
and at the LHC $\sqrt{s} = 14 TeV$ , respectively.}
\label{fig:sigma-g}
\end{center}
\end{figure}

In this scenario, to simplify the calculation we assumed 
$m_S=m_{\tilde Q}= m_{\tilde U}= m_{\tilde D}
=\mu =A_t=A_b=1~TeV$. As shown in Fig.~\ref{fig:sigma-g},
the SUSY-QCD decouples. The reason is 
the couplings $\alpha_{ij}$ in the vertex $H^{-}\tilde{t}_{i}\tilde{b}_{j}$ 
are fixed, and in this case the scalar
function like $C^6_0$ is proportional to  
$\frac{1}{m_{\tilde g}^2}\log \frac{m_S^2}{m_{\tilde g}^2}
\left[1+\frac{\hat{s}}{2m_{\tilde g}^2}\right]$ 
when $m^2_{\tilde{g}}\gg \hat{s}$ (see Eq.~(\ref{c0a},\ref{d0b})),
thus the SUSY QCD correction goes like 
$\frac{1}{m_{\tilde{g}}}\log\frac{m_S^2}{m_{\tilde g}^2}$.  

In the scenario, since $\hat{s}$ can be up to the large collider beam 
energy squared $s$, therefore,  not only the logarithmic dependence on 
the large mass parameter but also the large collider beam energies are 
responsible for the slow decoupling of the gluino, especially, at LHC, 
as shown in the figure. This is different from the previous studies 
in various decays\cite{hbb,hpt,Dobado02}.

(4) {\em Scenario D:} ~~Only squark masses are of the same order
(collectively denoted by $m_S$) and very large than 
other SUSY parameters and  the electroweak scale

\begin{figure}[hnt]
\begin{center}
\epsfig{file=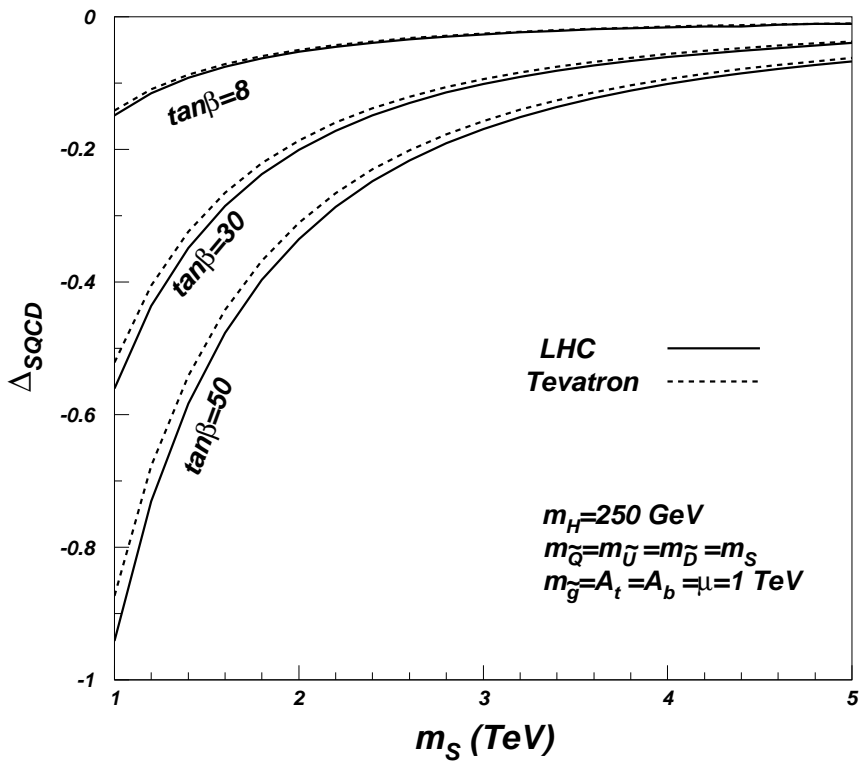,width=12cm}
\caption{Behavior of $\Delta_{SQCD}$ in the large squark masses
limit with fixed $m_{\tilde g} = A_b = A_t = \mu
= 1$ TeV and for different values of $\tan \beta$.The solid and dashed  
lines correspond to corrections at the Tevatron with 
$\sqrt{s} = 2 TeV$ and at the LHC $\sqrt{s} = 14 TeV$ , respectively.} 
\label{fig:sigma-ms}
\end{center}
\end{figure}

As shown in Fig.~\ref{fig:sigma-ms}, the SUSY QCD also decouples.  
In this case, it decouples much faster than in {\em Scenario C} where 
only gluino mass gets large. This can be understood easily because  
in this case the couplings $\alpha_{ij}$ in the vertex 
$H^{-}\tilde{t}_{i}\tilde{b}_{j}$ and the mass of gluino are both
fixed, so that the SUSY QCD correction goes like
$\sim C^{1}_{11}\to\frac{1}{m_S^2}\log\frac{m^2_{\tilde g}}{m_S^2}$.

Some remarks on the SUSY QCD correction to  the process of 
$gb \rightarrow tH^{\pm}$ are in order:

\begin{itemize}
\item[\rm 1.]
From the above analysis we find that the fundamental reason for such 
non-decoupling behavior of SUSY QCD
in the process  $gb \rightarrow tH^{\pm}$ is that some couplings like 
$H^{-}\tilde{t}_{i}\tilde{b}_{j}$ are proportional to SUSY mass parameters. 
This is similar to the non-decoupling property of the heavy top quark 
in the SM, where the top quark Yukawa couplings are proportional to top 
quark mass.  
\item[\rm 2.] 
 The non-decoupling behavior shown in Fig.~\ref{fig:sigma-gq}
 is in agreement with previous studies of SUSY-QCD corrections 
 in some decay process\cite{hbb,hpt}. In particular, the correction 
 $\Delta_{SQCD}$ shown in this figure as a function of the common 
 scale $m_S$ and $\tan\beta$ looks quite similar to the corresponding 
 corrections in the partial decay width $\Gamma (H^+\rightarrow t\bar b)$, 
 as given in Fig.2 of Ref.\cite{hbb}. The same kind of similarities are 
 also found between Fig.\ref{fig:sigma-g} and Fig. 6 of Ref.~\cite{hbb}, 
 and between  Fig.\ref{fig:sigma-ms} and Fig.7 of Ref.~\cite{hbb}.        
 Although the processes are different, the fundamental reason for such 
 non-decoupling behavior is the same.

\item[\rm 3.]
From  Figs.~\ref{fig:sigma-gq}-\ref{fig:sigma-ms} one sees that the size 
of SUSY-QCD corrections can be quite large for large $\tan\beta$. Note 
that when one-loop effects are too large, higher-level loops must be also 
calculated. We refer the reader to Ref.~\cite{CarenaDavid} where some 
techniques of resummation for a better convergence are proposed. 

\item[\rm 4.]
  As shown in Ref.~\cite{zhu01}, the genuine QCD corrections to this 
  process is also sizable, which can enhance the production rate 
  by $40\% \sim 80\%$ when charged Higgs mass and $\tan\beta$ vary
  in the range $180 \sim 1000$ GeV and $2 \sim 50$, respectively.  
  It is clear that the SUSY-QCD correction evaluated in this work 
  is comparable in size to the genuine QCD corrections. 
  It is noticeable that the genuine QCD corrections are always positive 
  whereas the SUSY-QCD corrections are negative in most SUSY parameter 
  space.  
\end{itemize}

\section{Conclusions}                                                    
\label{sec:conclusions}

In this work we have evaluated SUSY QCD radiative corrections to 
the $gb \rightarrow tH^{\pm}$ at
Tevatron and LHC. We have found that in some parameter space
the one-loop SUSY-QCD correction can be quite large and 
cannot be neglected. We have discussed in detail on 
the decoupling behavior of the corrections in the
large SUSY mass limit, and found that with fixed gluino mass
the one-loop SUSY QCD corrections decouple; while
non-decoupling occurs when gluino mass and $\mu$ or $A$ 
parameters both get large. The non-decoupling behavior of 
the SUSY-QCD corrections in the  process $gb\rightarrow t H^-$ 
is similar to the ones found in 
the literatures for the Higgs particles and top quark decays. 
We pointed out that such non-decoupling behavior 
arises from the $H^{-}\tilde{t}_{i}\tilde{b}_{j}$ vertices
which are proportional to SUSY mass parameters, 
as stated in some previous literatures\cite{hbb,hpt,Dobado02}.
Such large non-decoupling effects may play an important role
in the indirect search for SUSY from the production of
a top quark associated with a charged Higgs boson at 
Tevatron and LHC.


\section*{Acknowledgment}
This work is supported in part by a grant of Chinese Academy of Science
for Outstanding Young Scholars.


\section{Appendix A}
Here we list the coefficients $C^l$, scalar functions
$h_n^{(l)}$ and the vertex $V(H^{-}\tilde{t}_{i}\tilde{b}_{j})=
ig\alpha_{ij}/\sqrt{2}m_W ~(i,j = 1, 2)$ needed in our calculations. 

\begin{itemize}
\item Coefficients $C^l$
\end{itemize}

\begin{eqnarray}
C^{V(s)} &=& \frac{1}{\hat{s}-m_{b}^{2}},   
C^{V(t)} = \frac{1}{\hat{t}-m_{t}^{2}},\nonumber\\
C^{s(s)} &=& \frac{1}{(\hat{s}-m_{b}^{2})^{2}},
C^{s(t)} = \frac{1}{(\hat{t}-m_{t}^{2})^{2}}, ~C^{box} = 1.
\end{eqnarray}

\begin{itemize}
\item Scalar functions $h_{n}^{(l)}$
\end{itemize}

\begin{eqnarray}
h_{1}^{(l)} &=& 4m_{t}\eta_t(2p_{b}\cdot k -p^{(l)}\cdot p_b)
-4m_{b}\eta_b(p^{(l)}\cdot p_{t} +p_{t}\cdot k),\nonumber\\ 
h_{3}^{(l)} &=& 2\eta_t(2p_{b}\cdot kp_{b}\cdot p_{t}
-m_{b}^{2}p_{t}\cdot k -2p^{(l)}\cdot p_bp_{b}\cdot p_{t})
+2m_{b}m_{t}\eta_b(p_{b}\cdot k-2p^{(l)}\cdot p_b),\nonumber\\ 
h_{5}^{(l)} &=& 2\eta_t(m_{t}^{2}p_{b}\cdot k
-2p^{(l)}\cdot p_tp_{b}\cdot p_{t}) 
 +2m_{b}m_{t}\eta_b(p_{t}\cdot k -2p^{(l)}\cdot p_t),\nonumber\\ 
h_{7}^{(l)} &=& 4\eta_t(p^{(l)}\cdot
p_bp_{t}\cdot k -p^{(l)}\cdot kp_{b}\cdot p_{t} -p_b\cdot 
kp^{(l)}\cdot p_t -2p_{b}\cdot kp_{t}\cdot k)-4m_{b}m_{t}\eta_b p^{(l)}\cdot k,\nonumber\\ 
 h_{9}^{(l)} &=& 4m_{t}\eta_t p_{b}\cdot k(p_{b}\cdot k-p^{(l)}\cdot p_b)
 -4m_{b}\eta_bp^{(l)}\cdot p_bp_{t}\cdot k,\nonumber\\ 
h_{11}^{(l)} &=& 4m_{t}\eta_t p_{b}\cdot k(p_{t}\cdot k-p^{(l)}\cdot p_t)
-4m_{b}\eta_b p_{t}\cdot kp^{(l)}\cdot p_t,\nonumber\\ 
h_{2,4,6,8,10,12}^{(l)} &=&h_{1,3,5,7,9,11}^{(l)}(\eta_b\leftrightarrow\eta_t),
\end{eqnarray}
where the index $l$ represents the two channels $s$ and $t$, and
$p^{(s)}=p_b$, $p^{(t)}=p_t$.

\begin{itemize}
\item Couplings of $H^{-}\tilde{t}_{i}\tilde{b}_{j}$ 
\end{itemize}
The $H^{-}\tilde{t}_{i}\tilde{b}_{j}$ interaction terms can be
parametrized s 
\begin{eqnarray}
{\cal L}_{H^-\tilde{t}_{i}\tilde{b}_{j}}=
\frac{g}{\sqrt{2}m_W} \alpha_{ij} ~(i,j=1,2),
\label{coupling}
\end{eqnarray}
where 
\begin{eqnarray}
\alpha_{ij} = R_{i1}^{t*}R_{j1}^bg_{LL}
              + R_{i2}^{t*}R_{j2}^bg_{RR}
            + R_{i1}^{t*}R_{j2}^bg_{LR}
              + R_{i2}^{t*}R_{j1}^bg_{RL}
\label{coupling1}
\end{eqnarray}               
with             
\begin{eqnarray}
g_{LL} &=&-m_{W}^2\sin2\beta+m_{b}^2\tan\beta
+m_{t}^2\cot\beta,\nonumber\\ 
g_{RR} &=& m_{b}m_{t}(\tan\beta+\cot\beta),\nonumber\\ 
g_{LR} &=& m_{b}(\mu+A_{b}\tan\beta),\nonumber\\ 
g_{RL} &=& m_{t}(\mu+A_{t}\cot\beta).
\label{coupling2}
\end{eqnarray}  

\section*{Appendix B}

The form factors $F^{l}_{n}$ raise from 
the renormalized vertices and  propagators of s-channel and t-channel,
as well as box diagram given as following.
\begin{itemize}
\item{The renormalized vertices of s-channel:}
\end{itemize}
\begin{eqnarray}
 F_{1}^{V_{1}(s)}&=&
-3\eta_b\left\{A_{bi}^1m_{\tilde{g}}p_{b}\cdot kC^1_{0}+
A_{bi}^2m_{b}\left[\hat{s}C^1_{0}+m_{b}^2(C^1_{0}+4C^1_{11}+2C^1_{21})
\right.\right.\nonumber\\
&&\left.\left.+2p_{b}\cdot k(C^1_{11}+2C^1_{12}+2C^1_{23})
-2m_{\tilde{g}}^2C^1_{0}-1+4C^1_{24}
+m_{b}p_{b}\cdot k(C^1_{0}+C^1_{11})\right]\right\}
\nonumber\\
&&+\frac{4}{3}A_{bi}^2\eta_bm_{b}C^2_{24}
-\frac{16\pi^2}{g_{s}^2}\eta_tm_{b}
(\delta Z_R^{b}-\delta Z_L^{b}),\nonumber\\
F_{3}^{V_{1}(s)}&=&\frac{2}{3}\eta_t\left\{-
\left[m_{b}^2(C^1_{11}+C^1_{21})+p_{b}\cdot k(C^1_{12}+C^1_{23})
+C^1_{24}\right]+A_{bi}^1m_{\tilde{g}}m_{b}(C^1_{0}+C^1_{11})\right.\nonumber\\
&&\left.+2A_{bi}^2\left[C^1_{24}+p_{b}\cdot k(C^1_{12}+C^1_{23})\right]\right\}
-3\eta_t\left\{\left[m_{b}^2(C^2_{0}+2C^2_{11}+C^2_{21})+m_{\tilde{g}}^2C^2_{0}
-2C^2_{24}+1/2\right]\right.\nonumber\\ 
& &\left.+2A_{bi}^1m_{b}m_{\tilde{g}}(C^2_{0}+C^2_{11})
+2A_{bi}^2\left[m_{b}^2(C^2_{0}+2C^2_{11}+C^2_{21})
-m_{\tilde{g}}^2C^2_{0}-1/2+2C^2_{24}\right]\right\}\nonumber\\
&&+\frac{16\pi^2}{g_{s}^2}\eta_t2\delta Z_{R}^{b},\nonumber\\
F_{7}^{V_{1}(s)} &=& \frac{1}{3}\eta_t(C^1_{24}-2A_{bi}^2C^1_{24})
-\frac{3}{2}\eta_t\left\{m_{b}^2(C^2_{21}-C^2_{0})
+2p_{b}\cdot k(C^2_{12}+C^2_{23})-m_{\tilde{g}}^2C^2_{0}+2C^2_{24}-1/2 
\right.\nonumber\\ 
& &\left.-2A_{bi}^1m_{b}m_{\tilde{g}}C^2_{0}
-2A_{bi}^2\left[m_{b}^2(C^2_{0}+2C^2_{11}+C^2_{21})+2p_{b}\cdot k(C^2_{12}+C^2_{23})
-m_{\tilde{g}}^2C^2_{0}-1/2+2C^2_{24}\right]\right\}\nonumber\\ 
& &+\frac{16\pi^2}{g_{s}^2}\eta_t\delta Z_{R}^{b},\nonumber\\
F_{2,4,8}^{V_{1}(s)}  &=&  F_{1,3,7}^{V_{1}(s)}
\left(\eta_t\leftrightarrow \eta_b,
A_{bi}^2\rightarrow -A_{bi}^2, R\leftrightarrow L\right), \nonumber\\ 
F_{9}^{V_{1}(s)}&=& \frac{1}{3}\eta_b\left[-m_{b}(C^1_{11}+C^1_{21})
+A_{bi}^1m_{\tilde{g}}(C^1_{0}+C^1_{11})
+2A_{bi}^2m_{b}(2C^1_{12} +2C^1_{23}-C^1_{11}
-C^1_{21})\right]\nonumber\\ 
&&-3\eta_b\left[m_{b}(C^2_{11}+C^2_{21}) 
+A_{bi}^1m_{\tilde{g}}C^2_{11}+2A_{bi}^2m_{b}(C^2_{11}
+C^2_{21}-2C^2_{12}-2C^2_{23})\right],\nonumber\\
F_{10}^{V_{1}(s)} &=& F_{9}^{V_{1}(s)}\left(\eta_b\rightarrow \eta_t,
A_{bi}^2\rightarrow -A_{bi}^2\right),
\label{v1s}
\end{eqnarray}
where the Feyman integrals are defined as 
$C^1\equiv C\left(p_{b},k,m_{\tilde{b}_i}, 
m_{\tilde{g}},m_{\tilde{g}}\right)$, $C^2\equiv
C\left(-p_{b},-k,m_{\tilde{g}},m_{\tilde{b}_i},m_{\tilde{b}_i}\right)$,
$C^3\equiv C\left(-p_{t},-p_{H^-},
m_{\tilde{g}},m_{\tilde{t}_i},m_{\tilde{b}_j}\right)$
and
\begin{eqnarray}
F_{1}^{V_{2}(s)} &=& \frac{8}{3}\alpha_{ij}A_{ij}^{1R}
p_{b}\cdot kC^3_{12},\nonumber\\
F_{2}^{V_{2}(s)} &=& F_{1}^{V_{2}(s)}\left(R\rightarrow L\right),
\nonumber\\
F_{3}^{V_{2}(s)} &=& \frac{8}{3}\alpha_{ij}
\left[A_{ij}^{1L}m_{b}C^3_{12}+A_{ij}^{1R}m_{t}(C^3_{11}-C^3_{12})
-m_{\tilde{g}}A_{ij}^{2L}C^3_{0}\right]
-\frac{16\pi^2}{g_{s}^2}\eta_t\left[\delta Z_{R}^{t}+
\delta Z_{L}^{b}+2\frac{\delta m_t}{m_t}\right],\nonumber\\
F_{7}^{V_{2}(s)} &=& -\frac{4}{3}\alpha_{ij}
\left[A_{ij}^{1L}m_{b}C^3_{12}+A_{ij}^{1R}m_{t}(C^3_{11}-C^3_{12})
-A_{ij}^{2L}m_{\tilde{g}}C^3_{0}\right]
+\frac{16\pi^2}{g_{s}^2}\eta_t\left[\frac{1}{2}\delta Z_{R}^{t}+
\frac{1}{2}\delta Z_{L}^{b}+\frac{\delta m_t}{m_t}\right],\nonumber\\ 
F_{4,8}^{V_{2}(s)} &=& F_{3,7}^{V_{2}(s)}
\left(R\leftrightarrow L, \eta_t\rightarrow \eta_b,
\frac{\delta m_t}{m_t}\rightarrow\frac{\delta m_b}{m_b}\right).
\label{v2s}
\end{eqnarray}
Here $A_{(t,b)i}^1=(-1)^i\sin2\theta_{t,b}$, $A_{(t,b)i}^2
=\frac{1}{2}(-1)^i\cos 2\theta_{t,b}$. 
$A_{ij}$ are defined as $A_{ij}^{1L}=2R^t_{i2}R^b_{j2}$,
$A_{ij}^{1R}=2R^t_{i1}R^b_{j1}$, $A_{ij}^{2L}=-2R^t_{i2}R^b_{j1}$
and $A_{ij}^{2R}=-2R^t_{i1}R^b_{j2}$.

\begin{itemize}
\item{The renormalized vertices of t-channel}:
\end{itemize}

\begin{eqnarray}
F_{1}^{V_{1}(t)}  &=& \frac{4}{3}A_{ti}^2\eta_tm_{t}C^4_{24}
+3\eta_t\left\{m_{t}p_{t}\cdot k(C^5_{0}+C^5_{11})
+A_{ti}^1m_{\tilde{g}}p_{t}\cdot kC^5_{0}\right.\nonumber\\ 
& &\left.-A_{ti}^2m_{t}\left[\hat{t}C^5_{0}
+m_{t}^2(C^5_{0}+4C^5_{11}+2C^5_{21})
-2p_{t}\cdot k(C^5_{11}+2C^5_{12}+2C^5_{23}) 
-2m_{\tilde{g}}^2C^5_{0}-1+4C^5_{24})\right]\right\}\nonumber\\ 
& &-\frac{16\pi^2}{g_{s}^2}\eta_tm_{t}
\left(\delta Z_{R}^{t}-\delta Z_{L}^{t}\right),\nonumber\\ 
F_{5}^{V_{1}(t)}  &=& \frac{2}{3}\eta_t\left\{-
\left[m_{t}^2(C^4_{11}+C^4_{21})-p_{t}\cdot k(C^4_{12}+C^4_{23})
+C^4_{24}\right]+A_{ti}^1m_{\tilde{g}}m_{t}(C^4_{0}+C^4_{11})
\right.\nonumber\\ 
&&\left.+2A_{ti}^2\left[-C^4_{24}+p_{t}\cdot k(C^4_{12}+C^4_{23})\right]
\right\}
-3\eta_t\left\{m_{t}^2(C^5_{0}+2C^5_{11}+C^5_{21})
+m_{\tilde{g}}^2C^5_{0}-2C^5_{24}+1/2\right.\nonumber\\ 
& &\left.+2A_{ti}^1m_{t}m_{\tilde{g}}(C^5_{0}+C^5_{11})
-2A_{ti}^2\left[m_{t}^2(C^5_{0}+2C^5_{11}+C^5_{21})-m_{\tilde{g}}^2C^5_{0}
-1/2+2C^5_{24}\right]\right\}\nonumber\\ 
&&-\frac{16\pi^2}{g_{s}^2}\eta_t 2\delta Z_{R}^{t},\nonumber\\
F_{7}^{V_{1}(t)}&=& \frac{1}{3}\eta_t(C^4_{24}+2A_{ti}^2C^4_{24})
+\frac{3}{2}\eta_t\left\{m_{t}^2(C^5_{0}-C^5_{21})
+2p_{t}\cdot k(C^5_{12}+C^5_{23})+m_{\tilde{g}}^2C^5_{0}-2C^5_{24}+1/2
\right.\nonumber\\ 
& &\left.+2A_{ti}^1m_{t}m_{\tilde{g}}C^5_{0}
-2A_{ti}^2\left[m_{t}^2(C^5_{0}+2C^5_{11}
+C^5_{21})-2p_{t}\cdot k(C^5_{12}+C^5_{23})
-m_{\tilde{g}}^2C^5_{0}-1/2+2C^5_{24}\right]\right\}\nonumber\\ 
& &
+\frac{16\pi^2}{g_{s}^2}\eta_t\delta Z_{R}^{t},\nonumber\\  
 F_{2,6,8}^{V_{1}(t)}  &=&  F_{1,5,7}^{V_{1}(t)}\left(\eta_t\rightarrow \eta_b,
A_{ti}^2\rightarrow -A_{ti}^2, R\leftrightarrow L\right),
\nonumber\\
F_{11}^{V_{1}(t)} &=& \frac{1}{3}\eta_t\left[m_{t}(C^4_{11}+C^4_{21})
-A_{ti}^1m_{\tilde{g}}(C^4_{0}+C^4_{11})
-2A_{ti}^2m_{t}(2C^4_{12}+2C^4_{23}-C^4_{11} -C^4_{21})\right]\nonumber\\ 
& &+3\eta_t\left[m_{t}(C^5_{11}+C^5_{21})
+A_{ti}^1 m_{\tilde{g}}C^5_{11}
+A_{ti}^22m_{t}(C^5_{11}+C^5_{21}-2C^5_{12}-2C^5_{23})\right],
\nonumber\\ 
F_{12}^{V_{1}(t)} &=& F_{11}^{V_{1}(t)}\left(\eta_t\rightarrow \eta_b,
A_{ti}^2\rightarrow -A_{bi}^2\right),
\label{v1t}
\end{eqnarray}
where the Feyman integrals are defined as 
$C^4\equiv C\left(-p_{t},k,m_{\tilde{g}},
m_{\tilde{t}_i},m_{\tilde{t}_i}\right)$, $C^5\equiv
C\left(-p_{t},k,m_{\tilde{t}_i},m_{\tilde{g}},m_{\tilde{g}}\right)$, 
$C^6\equiv C\left(-p_{b},p_{H^-},
m_{\tilde{g}},m_{\tilde{b}_j},m_{\tilde{t}_i}\right)$ and 
\begin{eqnarray}
F_{1}^{V_{2}(t)} &=& -\frac{8}{3}\alpha_{ij}A_{ij}^{1R}
p_{t}\cdot kC^6_{12}\nonumber\\ 
F_{2}^{V_{2}(t)} &=&  F_{1}^{V_{2}(t)}\left(R\rightarrow L\right)
,\nonumber\\
F_{5}^{V_{2}(t)} &=&  \frac{8}{3}\alpha_{ij}\left[
A_{ij}^{1L}m_{b}(C^6_{11}-C^6_{12})
+A_{ij}^{1R}m_{t}C^6_{12}-A_{ij}^{2L}m_{\tilde{g}}C^6_{0}\right]
-\frac{16\pi^2}{g_{s}^2}\eta_t\left[\delta Z_{R}^{t}+
\delta Z_{L}^{b}+2\frac{\delta m_t}{m_t}\right],\nonumber\\
F_{7}^{V_{2}(t)} &=& -\frac{4}{3}\alpha_{ij}
\left[A_{ij}^{1L}m_{b}(C^6_{11}-C^6_{12})
+A_{ij}^{1R}m_{t}C^6_{12}
-A_{ij}^{2L}m_{\tilde{g}}C^6_{0}\right]
+\frac{16\pi^2}{g_{s}^2}\eta_t\left[\frac{1}{2}\delta Z_{R}^{t}+
\frac{1}{2}\delta Z_{L}^{b}+\frac{\delta m_t}{m_t}\right],\nonumber\\ 
 F_{6,8}^{V_{2}(t)} &=& F_{5,7}^{V_{2}(t)}
\left(R\leftrightarrow L, \eta_t\rightarrow \eta_b,
\frac{\delta m_t}{m_t}\rightarrow\frac{\delta m_b}{m_b}\right).
\label{v2t}
\end{eqnarray}

\begin{itemize}
\item{The renormalized propagator of s-channel}:
\end{itemize}
\begin{eqnarray}
F_{1}^{s(s)} &=& \frac{16\pi^2}{g_{s}^2}\eta_b
\left[2m_{b}p_{b}\cdot k
(\hat{\Sigma}_{L}^{b}+\hat{\Sigma}_{S}^{b})\right],\nonumber\\ 
F_{3}^{s(s)} &=& \frac{16\pi^2}{g_{s}^2}\eta_t\left[2\left(\hat{s}
\hat{\Sigma}_{L}^{b}+m_{b}^2\hat{\Sigma}_{R}^{b}\right)
+4m_{b}^2\hat{\Sigma}_{S}^{b}\right],\nonumber\\
 F_{7}^{s(s)} &=& -\frac{16\pi^2}{g_{s}^2}\eta_t\left[\hat{s}
\hat{\Sigma}_{L}^{b}+m_{b}^2\hat{\Sigma}_{R}^{b}
+2m_{b}^2\hat{\Sigma}_{S}^{b}\right],\nonumber\\
 F_{2,4,8}^{s(s)} &=& F_{1,3,7}^{s(s)}\left(\eta_t\leftrightarrow \eta_b,
R\leftrightarrow L\right).
\end{eqnarray}

\begin{itemize}
\item{The renormalized propagator of t-channel}:
\end{itemize}

\begin{eqnarray}
F_{1}^{s(t)} &=& -\frac{16\pi^2}{g_{s}^2}\eta_t
\left[2m_{t}p_{t}\cdot k
\left(\hat{\Sigma}_{L}^{t}+\hat{\Sigma}_{S}^{t}\right)\right],\nonumber\\ 
F_{5}^{s(t)} &=& \frac{16\pi^2}{g_{s}^2}\eta_t\left[2\left(\hat{t}
\hat{\Sigma}_{R}^{t}+m_{t}^2\hat{\Sigma}_{L}^{t}\right)
+4m_{t}^2\hat{\Sigma}_{S}^{t}\right], \nonumber\\ 
 F_{7}^{s(t)} &=& -\frac{16\pi^2}{g_{s}^2}\eta_t\left[\hat{t}
\hat{\Sigma}_{R}^{t}+m_{t}^2\hat{\Sigma}_{L}^{t}
+2m_{t}^2\hat{\Sigma}_{S}^{t}\right],
\nonumber\\ 
F_{2,6,8}^{s(t)} &=& F_{1,5,7}^{s(t)}\left(\eta_t\rightarrow \eta_b,
R\leftrightarrow L\right).
\end{eqnarray}

\begin{itemize}
\item{Box diagram contribution}:
\end{itemize}

\begin{eqnarray}
F_{1}^{box}  &=&\frac{1}{3}\alpha_{ij}A_{ij}^{1R}
\left\{9\left[D^1_{27}+\frac{1}{2}(m_{t}^{2}D^1_{22}+m_H^2D^1_{23})
-p_{t}\cdot kD^1_{24} +(p_{t}\cdot k-p_{b}\cdot k)D^1_{25}
\right.\right.\nonumber\\
&&\left.\left.+(p_{t}\cdot p_{b}+p_{t}\cdot k-m_{t}^{2})D^1_{26}
-p_{t}\cdot k(D^1_{12}-D^1_{13})
-\frac{1}{2}m_{\tilde{g}}^{2}D^1_{0}\right]
-D^2_{27}\right\}, \nonumber\\ 
 F_{3}^{Box}  &=& \frac{1}{3}\alpha_{ij}
\left\{-9\left[A_{ij}^{1L}m_{b}D^1_{23}+A_{ij}^{1R}m_{t}(D^1_{23}-D^1_{26})
-A_{ij}^{2L}m_{\tilde{g}}D^1_{13}\right]\right. \nonumber\\ 
&&\left.+A_{ij}^{1L}m_{b}(D^2_{13}-D^2_{11}-D^2_{21}-D^2_{23}+2D^2_{25})
-A_{ij}^{1R}m_{t}(D^2_{13}+D^2_{25}-D^2_{23})\right.\nonumber\\
 & &\left.+A_{ij}^{2L}m_{\tilde{g}}(D^2_{0}+D^2_{11}
-D^2_{13}) \right\}, \nonumber\\
 F_{5}^{box}  &=& \frac{1}{3}\alpha_{ij}
\left\{9\left[A_{ij}^{1L}m_{b}(D^1_{23}-D^1_{26}) 
-A_{ij}^{1R}m_{t}(D^1_{22}+D^1_{23}-2D^1_{26})
+A_{ij}^{2L}m_{\tilde{g}}(D^1_{12}-D^1_{13})\right] \right. \nonumber\\ 
&&\left.+A_{ij}^{1L}m_{b}(D^2_{23}-D^2_{25})-A_{ij}^{1R}m_{t}D^2_{23}
+A_{ij}^{2L}m_{\tilde{g}}D^2_{13}\right\},\nonumber\\ 
F_{7}^{box}  &=&-\frac{3}{2}\alpha_{ij}\left[
A_{ij}^{1L}m_{b}D^1_{13}+A_{ij}^{1R}m_{t}(D^1_{12}-D^1_{13})
-A_{ij}^{2L}m_{\tilde{g}}D^1_{0}\right], \nonumber\\
F_{9}^{box}  &=& \frac{1}{3}\alpha_{ij}A_{ij}^{1R}\left\{-9 
(D^1_{23}-D^1_{25})+D^2_{13}-D^2_{12}-D^2_{24} 
-D^2_{23}+D^2_{25} + D^2_{26}\right\}, \nonumber\\ 
F_{11}^{box} &=& \frac{1}{3}\alpha_{ij}A_{ij}^{1R}
\left\{9\left[D^1_{12}-D^1_{13}+D^1_{23}+D^1_{24}-D^1_{25}-D^1_{26}\right]
D^2_{23}-D^2_{26}\right\}, \nonumber\\
 F_{2,4,6,8,10,12}^{box} &=& F_{1,3,5,7,9,11}^{box}
\left(R\leftrightarrow L\right) 
\label{vbox}
\end{eqnarray}
where four-point functions are defined as $D^1\equiv
D\left(k,-p_{t},-p_{H^-},m_{\tilde{g}}, m_{\tilde{g}},
m_{\tilde{t}_i},m_{\tilde{b}_j}\right)$, $D^2\equiv 
D\left(-p_{b},-k,p_{H^-},m_{\tilde{g}},m_{\tilde{b}_j},
m_{\tilde{b}_j},m_{\tilde{t}_i}\right)$. 
Note in the above formula we take the convention that repeated 
indices are summed over.

All other form factors $F_n^l$ not listed above vanish. 
Note that the contributions from diagrams 
Fig.~\ref{fig:feyman}(i),\ref{fig:feyman}(j) and 
\ref{fig:feyman}(k) just give the renormalization constant 
$\delta Z_2^g$, which is canceled out by $\delta g_s$
due to the renormalization condition in Eq. (\ref{gs}). 

The renormalization constants appearing in the above form factors 
are given by
\begin{eqnarray}
\delta Z_L^t&=&-\frac{g_s^2C_F}{16\pi^2}\left\{m_t
\left[2m_t\frac{\partial B_1}{\partial p_t^2}
-m_{\tilde{g}}A_{ti}^1\frac{\partial B_0}{\partial p_t^2}\right]
\left \vert_{p_t^2=m_t^2}+(1-2A_{ti}^2)B_1\right.\right\}
\left(p_t^2, m_{\tilde{g}}^2, m_{\tilde{t}_i}^2\right),\nonumber\\
\delta Z_L^b&=&-\frac{g_s^2C_F}{16\pi^2}\left\{
m_b\left[2m_b\frac{\partial B_1}{\partial p_b^2}
-m_{\tilde{g}}A_{bi}^1\frac{\partial B_0}{\partial p_b^2}
\right]\left \vert_{p_b^2=m_b^2}
+(1-2A_{bi}^2)B_1\right.\right\}
\left(p_b^2, m_{\tilde{g}}^2, m_{\tilde{b}_i}^2\right),\nonumber\\
\delta Z_R^t&=&\delta Z_L^t(A_{ti}^2\rightarrow -A_{ti}^2),\nonumber\\
\delta Z_R^b&=&\delta Z_L^b(A_{bi}^2\rightarrow -A_{bi}^2),\nonumber\\
\delta m_t/m_t&=&\frac{g_s^2C_F}{16\pi^2}(B_1-\frac{m_{\tilde{g}}}{m_t}
A_{ti}^1B_0)\left(p_t^2, m_{\tilde{g}}^2, m_{\tilde{t}_i}^2\right),\nonumber\\
\delta m_b/m_b&=&\frac{g_s^2C_F}{16\pi^2}(B_1-\frac{m_{\tilde{g}}}{m_b}
A_{bi}^1B_0)\left(p_b^2, m_{\tilde{g}}^2, m_{\tilde{b}_i}^2\right),
\label{vselfa}
\end{eqnarray} 
where the color factor $C_F=4/3$. The renormalized self energy 
contributions from quarks as follows
\begin{eqnarray}
\hat{\Sigma}_{S}^{t}&=&\frac{g_s^2C_F}{16\pi^2}
\frac{m_{\tilde{g}}}{m_t}A_{ti}^1
\left[B_0\left(m_t^2, m_{\tilde{g}}^2, m_{\tilde{t}_i}^2\right)
- B_0\left(\hat{t}, m_{\tilde{g}}^2, m_{\tilde{t}_i}^2\right)\right],
\nonumber\\ 
\hat{\Sigma}_{S}^{b}&=&\frac{g_s^2C_F}{16\pi^2}
\frac{m_{\tilde{g}}}{m_b}A_{bi}^1
\left[B_0\left(m_b^2, m_{\tilde{g}}^2, m_{\tilde{b}_i}^2\right)
- B_0\left(\hat{s}, m_{\tilde{g}}^2, m_{\tilde{b}_i}^2\right)\right],
\nonumber\\ 
\hat{\Sigma}_L^t&=&-\frac{g_s^2C_F}{16\pi^2}
(1-2A_{ti}^2)B_1
\left(\hat{t}, m_{\tilde{g}}^2, m_{\tilde{t}_i}^2\right)
-\delta Z_L^t,\nonumber\\ 
\hat{\Sigma}_{L}^{b}&=&-\frac{g_s^2C_F}{16\pi^2}
(1-2A_{ti}^2)B_1
\left(\hat{s}, m_{\tilde{g}}^2, m_{\tilde{b}_i}^2\right)
-\delta Z_{L}^{b},\nonumber\\ 
\hat{\Sigma}_{R}^{t}& =&\hat{\Sigma}_{L}^{t}(A_{ti}^2\rightarrow -A_{ti}^2)
,\nonumber\\ 
\hat{\Sigma}_{R}^{b}&=&\hat{\Sigma}_{L}^{b}(A_{bi}^2\rightarrow -A_{bi}^2).
\label{vselfb}
\end{eqnarray} 

\section*{Appendix C}
In this appendix we give the expansions of the scalar loop
integrals in the asymptotic large mass limit. The definitions and
conventions of the Feynman loop integral functions can be found in 
\cite{bcanaly}. The integrals are performed in $4-\epsilon$ dimension
and the divergent contributions are regularized by $\Delta\equiv (2/\epsilon)
-\gamma_E+\log(4\pi)-log(m_h^2/\mu^2_0)$ with $m_h$ being 
the corresponding mass of the heavy particle in the loops and 
$\mu_0$, the regularization scale.  

Under the assumption of $m_h^2=max(m_k^2)
\gg p_i\cdot p_j\ (i,j=1,2; k=1,2,3)$, 
we consider special cases used in our calculations. 
For the two-point function $B_1(p^2, m_h^2, m^2_l)$, we obtain
\begin{eqnarray}
B_1=\frac{1}{2}
\left[\Delta+\frac{3-\delta}{2(1-\delta)}+\frac{2\delta-\delta^2}
{(1-\delta)^2}\log\delta\right\}+{\cal O}\left(\frac{p^2}{m_h^2}\right)
\end{eqnarray}
where $\delta=m_l^2/m_h^2$. Therefore,  the asymptotic form can be 
expressed as
\begin{equation}
B_1=\left\{\begin{array}{ll}
\frac{\Delta}{2}+\frac{p^2}{12m_h^2}, & \delta\to 1,\\
\frac{\Delta}{2}+\frac{3}{4}+\frac{p^2}{3m_h^2}, & \delta\to 0,\\
\frac{\Delta}{2}+\frac{1}{4}+\frac{p^2}{6m_h^2}, 
& \delta\to \infty.
\end{array}
\right.
\label{b1}
\end{equation}
Note in the case $\delta\to\infty$, $m_h$ in Eq.~(\ref{b1}) should be 
replaced  by $m_l$.  

For the three-point function $C_{(0,24)}(p_1, p_2, m_1^2, m_2^2, m_3^2)$ 
we expand them as follows:

\begin{itemize} 
\item Case A: $m_h=m_1, m_l^2=m_2^2=m_3^2\pm\Delta_m^2$
\end{itemize}
\begin{eqnarray}
C_0&=&\frac{1}{m_h^2}\left\{\frac{1}{1-\delta}+
\frac{1}{(1-\delta)^2}\ln \delta+\frac{p^2}{m_h^2}\left[
\frac{\delta+5}{4(1-\delta)^3}+\frac{2\delta+1}{2(1-\delta)^4}
\ln \delta\right]
+{\cal O}\left(\frac{\Delta_m^2}{m_h^2}\right)
+{\cal O}\left(\frac{p^4}{m_h^4}\right)\right\},\nonumber\\
C_{24}&=&\Delta+\frac{3-\delta}{8(1-\delta)}
+\frac{2\delta-\delta^2}{4(1-\delta)^2}\ln \delta
+\frac{p^2}{24m_h^2}\left[
\frac{2+5\delta-\delta^2}{(1-\delta)^3}+\frac{6\delta}{(1-\delta)^4}
\ln \delta\right]+{\cal O}\left(\frac{\Delta_m^2}{m_h^2}\right)
+{\cal O}\left(\frac{p^4}{m_h^4}\right)
\label{c024}
\end{eqnarray}
where $p=p_1+p_2$. In asymptotic large mass limit
we obtain
\begin{equation}
C_0=\left\{\begin{array}{ll}
-\frac{1}{2m_h^2}\left[1
+\frac{p^2}{12m_h^2}+{\cal O}\left(\frac{\Delta_m^2}{m_h^2}\right)
+{\cal O}\left(\frac{p^4}{m_h^4}\right)\right], &\delta\to 1,\\
\frac{1}{m_h^2}\ln\frac{m_l^2}{m_h^2}\left[1+\frac{p^2}{2m_h^2}
+{\cal O}\left(\frac{\Delta_m^2}{m_h^2}\right)
+{\cal O}\left(\frac{p^4}{m_h^4}\right)\right], &\delta\to 0.
\end{array}
\right.
\label{c0a}
\end{equation}
 \begin{equation}
C_{24}=\left\{\begin{array}{ll}
\Delta+\frac{p^2}{48m_h^2}
+{\cal O}\left(\frac{\Delta_m^2}{m_h^2}\right)
+{\cal O}\left(\frac{p^4}{m_h^4}\right), &\delta\to 1,\\
\Delta+\frac{3}{8}+\frac{p^2}{12m_h^2}
+{\cal O}\left(\frac{\Delta_m^2}{m_h^2}\right)
+{\cal O}\left(\frac{p^4}{m_h^4}\right), &\delta\to 0.\\
\end{array}
\right.
\label{c24a}
\end{equation}
 
\begin{itemize} 
\item Case B: $m_l=m_1, m_h=m_2=m_3$
\end{itemize}
\begin{eqnarray}
C_0&=&-\frac{1}{m_h^2}\left\{\frac{1}{1-\delta}+
\frac{\delta}{(1-\delta)^2}\ln \delta+\frac{p^2}{m_h^2}\left[
\frac{1+5\delta}{4(1-\delta)^3}+\frac{\delta^2+2\delta}{2(1-\delta)^4}
\ln \delta\right] 
+{\cal O}\left(\frac{p^4}{m_h^4}\right)\right\},\nonumber\\
C_{24}&=&\Delta+\frac{1-3\delta}{8(1-\delta)}
-\frac{\delta^2}{4(1-\delta)^2}\ln \delta
+\frac{p^2}{24m_h^2}\left[
\frac{1-5\delta-2\delta^2}{(1-\delta)^3}-\frac{6\delta^2}{(1-\delta)^4}
\ln \delta\right]
+{\cal O}\left(\frac{p^4}{m_h^4}\right).
\end{eqnarray}
The asymptotic form is given by
\begin{equation}
C_0=\left\{\begin{array}{ll}
-\frac{1}{2m_h^2}\left[1+\frac{p^2}{12m_h^2}
+{\cal O}\left(\frac{p^4}{m_h^4}\right)\right], &\delta\to 1,\\
-\frac{1}{m_h^2}\left[1+\frac{p^2}{4m_h^2}
+{\cal O}\left(\frac{p^4}{m_h^4}\right)\right], &\delta\to 0,
\end{array}
\right.
\label{c0b}
\end{equation}
\begin{equation}
C_{24}=\left\{\begin{array}{ll}
\Delta+\frac{p^2}{48m_h^2}
+{\cal O}\left(\frac{p^4}{m_h^4}\right), &\delta\to 1,\\
\Delta+\frac{1}{8}+\frac{p^2}{24m_h^2}
+{\cal O}\left(\frac{p^4}{m_h^4}\right), &\delta\to 0.\\
\end{array}
\right.
\label{c24b}
\end{equation}

For the four-point functions $D_{0,1k}
(p_1,p_2,p_3, m_1^2, m_2^2, m_3^2, m_4^2)$ with 
the assumption of $m_h^2=max(m_k^2)\gg p^2=max(p_i\cdot p_j)
\ (i,j=1,2,3; k=1,2,3,4)$, 
we consider two special cases used in our calculations 

\begin{itemize} 
\item Case A: $m_h=m_1, m_l=m_2$
\end{itemize}
\begin{eqnarray}
D_0&=&\frac{1}{m_h^2m_l^2}\left\{\frac{1+\delta}{2(1-\delta)^2}+
\frac{\delta}{(1-\delta)^3}\ln \delta
+{\cal O}\left(\frac{\Delta^2}{m_h^2}\right)
+{\cal O}\left(\frac{p^2}{m_h^2}\right)\right\},\nonumber\\
D_{11}&=&\frac{1}{m_h^2m_l^2}\left\{\frac{2+5\delta-\delta^2}
{4(1-\delta)^3}+\frac{3\delta}{(1-\delta)^4}\ln \delta
+{\cal O}\left(\frac{\Delta_m^2}{m_h^2}\right)
+{\cal O}\left(\frac{p^2}{m_h^2}\right)\right\},\nonumber\\
D_{12}&=&\frac{2}{3}D_{11},~ D_{13}=\frac{1}{3}D_{11}
\end{eqnarray}
where $\Delta_m^2=max(|m_2^2-m_3^2|, |m_2^2-m_4^2|,|m_3^2-m_4^2|)$.
In asymptotic large mass limit we obtain
\begin{equation}
D_{(0,11,12,13)}=\left\{\begin{array}{ll}
\left(\frac{1}{6},\frac{1}{8},\frac{1}{12},\frac{1}{24}\right)
\frac{1}{m_h^4}\left[1
+{\cal O}\left(\frac{p^2}{m_h^2}\right)\right], &\delta\to 1,\\
\left(\frac{1}{2},\frac{1}{2},\frac{1}{3},\frac{1}{6}\right)
\frac{1}{m_h^2m_l^2}
\left[1+{\cal O}\left(\frac{p^2}{m_h^2}\right)\right], &\delta\to 0.
\end{array}
\right.
\label{d0a}
\end{equation}
 
\begin{itemize} 
\item Case B: $m_h=m_1=m_2, m_l^2=m_3^2=m_4^2\pm\Delta_m^2$
\end{itemize}
\begin{eqnarray}
D_0&=&-\frac{1}{m_h^4}\left\{\frac{2}{(1-\delta)^2}+
\frac{1+\delta}{(1-\delta)^3}\ln \delta
+{\cal O}\left(\frac{\Delta_m^2}{m_h^2}\right)
+{\cal O}\left(\frac{p^2}{m_h^2}\right)\right\},\nonumber\\
D_{11}&=&\frac{1}{m_h^4}\left\{\frac{9+3\delta}
{4(1-\delta)^3}+\frac{\delta^2-2\delta-2}{2(1-\delta)^4}\ln \delta
+{\cal O}\left(\frac{\Delta_m^2}{m_h^2}\right)
+{\cal O}\left(\frac{p^2}{m_h^2}\right)\right\},\nonumber\\
D_{12}&=&-\frac{1}{m_h^4}\left\{\frac{5+\delta}
{2(1-\delta)^3}+\frac{1+2\delta}{(1-\delta)^4}\ln \delta
+{\cal O}\left(\frac{\Delta_m^2}{m_h^2}\right)
+{\cal O}\left(\frac{p^2}{m_h^2}\right)\right\},\nonumber\\
D_{13}&=&\frac{1}{2}D_{12}.  
\end{eqnarray}
In asymptotic large mass limit we obtain
\begin{equation}
D_{(0,11,12,13)}=\left\{\begin{array}{ll}
\left(\frac{1}{6},\frac{1}{8},\frac{1}{12},\frac{1}{24}\right)
\frac{1}{m_h^4}
\left[1+{\cal O}\left(\frac{\Delta^2}{m_h^2}\right)
+{\cal O}\left(\frac{p^2}{m_h^2}\right)\right], &\delta\to 1,\\
\left(-1,-1,-1,-\frac{1}{2}\right)
\frac{1}{m_h^4}\ln\frac{m_l^2}{m_h^2}\left[1
+{\cal O}\left(\frac{\Delta_m^2}{m_h^2}\right)
+{\cal O}\left(\frac{p^2}{m_h^2}\right)\right], &\delta\to 0.
\end{array}
\right.
\label{d0b}
\end{equation}


\begingroup\raggedright\endgroup

\end{document}